\documentclass[twocolumn,aps,showkeys,preprintnumbers,amsmath,amssymb]{revtex4}

\usepackage{graphicx}
\usepackage{dcolumn}
\usepackage{bm}

\begin{document}


\title{Tunneling between Two Quantum Hall Droplets}

\author{Inseok Yang}
\author{Woowon Kang}%
\affiliation{%
James Franck Institute and Department of Physics\\
 University of Chicago, Chicago, Illinois 60637
}%

\author{Loren Pfeiffer}
 \author{Kirk Baldwin}
 \author{Ken West}
\affiliation{
Bell Laboratories, Lucent Technologies \\
600 Mountain Avenue, Murray Hill, NJ 07974
}%

\date{\today}

\begin{abstract}
We report on tunneling experiment between two quantum Hall droplets separated by a nearly ideal tunnel barrier. The device is produced by cleaved edge overgrowth that laterally juxtaposes two two-dimensional electron systems across a high quality semiconductor barrier. The dramatic evolution of the tunneling characteristics is consistent with the magnetic field-dependent tunneling between the coupled edge states of the quantum Hall droplets. We identify a series of quantum critical points between successive strong and weak tunneling regimes that are reminiscent of the plateau-transitions in quantum Hall effect.  Scaling analysis shows that the conductance near the critical magnetic fields $B_{c}$ is a function of a single scaling argument $|B-B_{c}|T^{-\kappa}$, where the exponent $\kappa = 0.42$. This puzzling resemblance to a quantum Hall-insulator transition points to the significance of interedge correlation in the lateral tunneling of quantum Hall droplets.
\end{abstract}

\maketitle

\section{\label{sec:level1} Introduction}

The edge states in the quantum Hall regime represents a nearly ideal realization of one-dimensional electronic systems in nature\cite{KaneFisher}. Analogous to the skipping orbits for a metallic sheet within a semi-classical picture, the edge state represents the only possible physical excitation for a quantum Hall droplet. Perhaps the most remarkable feature of the edge state involves its unique ability to adjust itself spatially as well as energetically associated with its gapless property. Resulting ballistic motion of the edge electrons precludes backscattering, making it an ideal system for studying the effect 
of interaction in one-dimensional electronic systems. In the integer quantum Hall regime, the edge state contains a discrete number of modes, one for each filled Landau level, that contributes a quantum of $e^{2}/h$ to the overall conductance. For a fractional quantum Hall state at $\nu = 1/m$, the edge mode consists of a single chiral Luttinger liquid whose Luttinger parameter $g$ specified by the filling factor\cite{Wen90a,Wen91}.

Following the prediction of chiral Luttinger liquids in the fractional quantum Hall effect\cite{Wen90a,Wen91}, extensive effort has been devoted to the study of tunneling between quantum Hall edge states\cite{Kane92,Kane94,Fendley95,Chamon97,Milliken96,Chang96,Grayson98,Shytov98}. 
Experimental studies of tunneling between edge states across a quantum  point contact\cite{Milliken96} and tunneling between an edge state and a three-dimensional metal\cite{Chang96,Grayson98} have generally tended to support the predicted Luttinger liquid behavior. However, there remain important open questions regarding the experimentally observed exponent and its correlation to the bulk quantum Hall states\cite{Shytov98}.

A different and perhaps more intriguing geometry for the study of edge state tunneling involves a line junction that juxtaposes two quantum Hall droplets separated by a well-defined tunnel barrier. Such a junction in the dirty limit has been initially envisioned as a Hall bar with a long narrow gate that couples two fractional quantum Hall liquids\cite{Renn95,Kane97}. The resulting junction produces two parallel, counterpropagating edge modes against each other across the barrier. The inter-mode backscattering from the defects in the barrier has led to prediction for a metal-insulator transition in the conductance across the junction\cite{Renn95,Kane97}. 
In the clean limit, a coupled Luttinger liquid emerges when the edge states of integer quantum Hall are placed close to each other\cite{Kim03}. Resonant mixing of the states with equal transverse momentum\cite{Ho94,Takagaki00,Nonoyama02} has led to prediction for a one-dimensional broken symmetry state\cite{Mitra01}.

In this paper, we report on the observation of a cascade of quantum phase transitions exhibited by tunnel-coupled edge states of quantum Hall line junctions. Two counterpropagating edge states are separated by an 8.8 nm-wide, $\sim$100$\mu$m-long semiconductor barrier. 
We identify a series of quantum critical points between successive strong and weak tunneling regimes that are reminiscent of the metal-insulator transition in two-dimensions. Scaling analysis shows that the conductance near the critical magnetic field $B_{c}$ is a function of a single scaling argument $|B-B_{c}|T^{-\kappa}$, where the exponent $\kappa \approx 0.42$. This apparent similarity to the quantum Hall-insulator transitions is quite puzzling due to
one-dimensional character of edge states. Whether the resemblance to a quantum Hall-insulator transition is coincidental or occurs from some deeper physics remains to be clarified.

\section{Experimental Details}

The line junctions which couple two quantum Hall liquids are fabricated by cleaved edge overgrowth using molecular beam epitaxy\cite{Pfeiffer90,Kang00,Yang04}. The initial growth on a standard (100) GaAs substrate consists of an undoped 13$\mu m$ GaAs layer followed by an  8.8 nm-thick digital alloy of undoped Al$_{0.1}$Ga$_{0.9}$As/AlAs, and completed by a 14$\mu m$ layer of undoped GaAs. This multilayer sample is cleaved along the (110) plane in an MBE machine and a modulation-doping sequence is performed over the exposed edge, forming two strips of two-dimensional electron systems separated from each other by the 8.8 nm-thick barrier. A mesa incorporating the barrier and the two-dimensional electron systems 
into a junction that is $\sim$100$\mu m$ long is defined by photolithography. The inset of Fig.~\ref{Fig:lnjnc}a shows the layout of the line junction device in the plane of the two-dimensional electrons. Under strong transverse magnetic field, Landau quantization creates two droplets of quantum Hall liquids separated by the rectangular tunnel barrier. The resulting junction places two counterpropagating edge states within the magnetic length of each other, leading to the geometry for inter-edge tunneling. The density of the two-dimensional electron system in the devices studied was $n = 2\times 10^{11} cm^{-2}$ with a mobility of $\sim 1\times 10^{5} cm^{2}/Vsec$. 

\section{Differential Conductance at Zero-Bias}

\begin{figure}
\includegraphics[width= \linewidth]{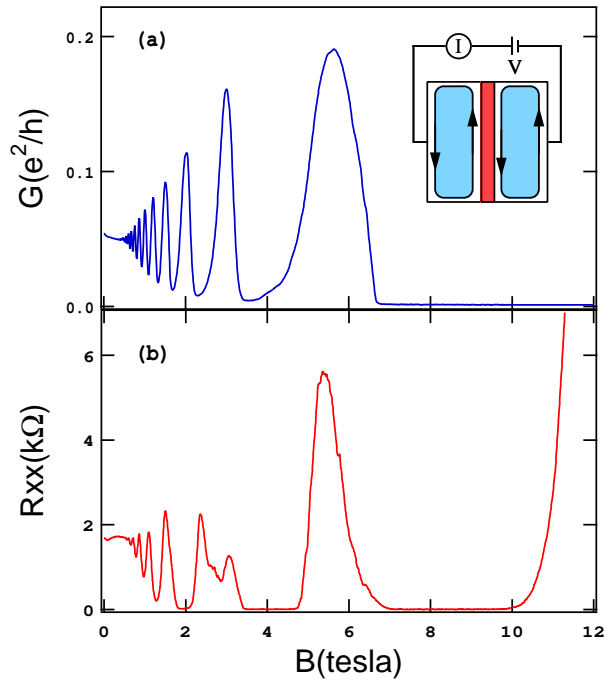}
\caption{\label{Fig:lnjnc} Top: Representative differential conductance, $G = dI/dV$, of the line junction at 300 mK. Bottom: Magnetoresistance from one of the two-dimensional electron system in the line junction. Inset: Layout of the line junction and measurement geometry. Two counterpropagating edge states are juxtaposed against the barrier in the quantum Hall regime.}
\end{figure}

Figure~\ref{Fig:lnjnc} illustrates the differential conductance at zero-bias, $G = dI/dV$, across the line junction and the magnetoresistance of the two-dimensional electron system
parallel to the tunnel barrier. The conductance exhibits a series of conductance peaks that oscillates with increasing magnetic field before abruptly dropping to zero above 6.7 tesla. No oscillatory features can be seen at higher magnetic fields. Shubnikov-de Haas oscillations are found in the magnetoresistance for low magnetic fields and integer quantum Hall states beyond 2 tesla. The period of Shubnikov-de Haas oscillations of the two-dimensional electron systems does not match the conductance oscillations, which are sharper and more distinct than the oscillations in the magnetoresistance. 

Two different theoretical scenarios have been proposed to describe the physics of the line junction in the clean limit. Within the level-mixing picture of tunneling across the line junction\cite{Ho94,Takagaki00,Nonoyama02,Mitra01,Kollar02}, the conductance peaks occur whenever the energy levels of the two edges coincide with the Fermi level at zero bias as a function of magnetic field. It was also proposed that the differential conductance peak arises from the formation of a correlated electronic state with spontaneous inter-edge coherence at zero momentum transfer\cite{Mitra01}. In an alternate picture, the differential conductance peak is due to the effects of point-contact tunneling in the Coulomb-coupled edge states\cite{Kim03}. In this framework the successive differential conductance peaks are due to quantum phase transitions tuned by the magnetic field, caused by opening and closing of tunneling channels  between the coupled edge states as the magnetic field is varied.  

\begin{figure}
\includegraphics[width= \linewidth]{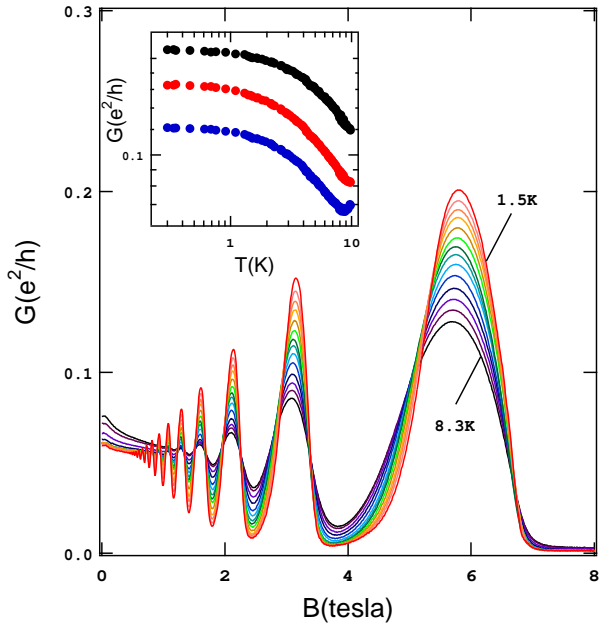}
\caption{\label{Fig:tdep} Conductance of a line junction for various temperature between 1.5K and 8.3K.
Inset: Temperature dependence of  conductance the first 3 peaks. }
\end{figure}

Figure~\ref{Fig:tdep} shows the magnetic field dependence of the zero bias conductance between 1.5 and 8.3K. The conductance peaks grow in amplitude with increasing magnetic field and decreasing temperature. Above 7 tesla, the differential conductance becomes vanishingly small as the 
momentum conserved tunneling across the line junction can no longer be satisfied and the conduction occurs parallel to the junction, along the barrier. A striking feature of the conductance data in Fig.~\ref{Fig:tdep} is the series of critical points on the high 
field side of the conductance peaks. These critical points separate 
the differential conductance peaks from the low conductance regions where the tunneling 
is largely suppressed. Interestingly, no critical points can be seen 
on the lower field side of the differential conductance peaks. In terms of single particle 
levels, there are excess states above the energy level crossings 
on the low field side prior to the entry into the zero-bias peaks. 
On the other hand, electronic states are depopulated as soon as 
the system exits the differential conductance peaks on the high field side. Consequently,
the observed aymmetry may 
be reflecting the structure of the energy level crossings as the 
population of the filled states change as a function of magnetic 
field. 
The inset of Fig.~\ref{Fig:tdep} illustrates the 
temperature dependence of differential conductance at the three largest zero 
bias-conductance peaks. Differential conductance increases slowly as temperature is 
reduced and saturates below 1K. 

\section{Quantum Critical Points and Scaling Analysis}  

\begin{figure}
\includegraphics[width= \linewidth]{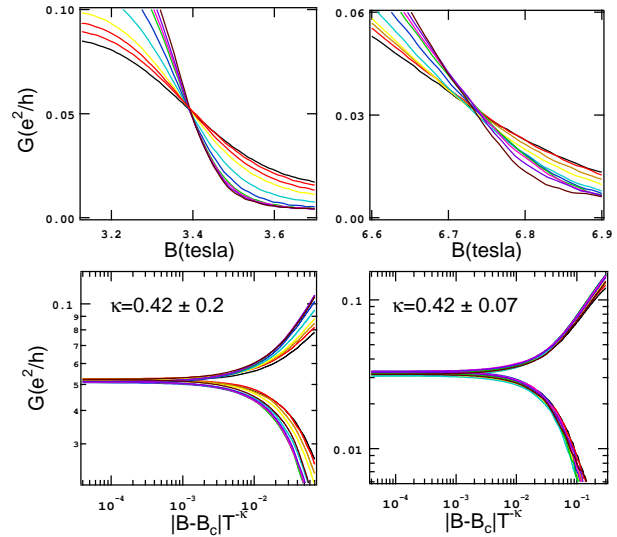}
\caption{\label{Fig:tran} Critical points and scaling analysis of the tunneling conductance. 
(a) Conductance data near $B_{c}$ = 3.39T. (b) Conductance data near 
$B_{c}$ = 6.73T. (c) Scaling analysis of the conductance data 
near $B_{c}$ = 3.39T as a function of $|B-B_{c}|T^{-\kappa}$. (d) 
Similar analysis performed for data near near $B_{c}$ = 6.73T.}
\end{figure}

Figures~\ref{Fig:tran}a and \ref{Fig:tran}b show an expanded view 
of the differential conductance near the critical points around $B_{c}$ = 3.39T and 6.73T 
with corresponding critical conductance values of $G_{c}$ of 
approximately $0.05e^{2}/h$ and $0.03e^{2}/h$. Immediately above (below) the critical magnetic field, $B_{c}$, differential conductance decreases (increases) with temperature. Such a behavior about the critical points is reminiscent of the quantum Hall-insulator 
transitions in two-dimensions. Figure~\ref{Fig:tran}c and \ref{Fig:tran}d 
illustrate the results of the scaling analysis of the differential conductance about 
the critical points.  For both cases we find that the tunneling conductance $G$ near $B_{c}$ can be scaled as an argument of $|B-B_{c}|T^{-\kappa}$, where $\kappa = 0.42$.
While the critical point at $B_{c}$ = 3.39T features a limited scaling regime and consequently a greater uncertainty in the value of the critical exponent, the extended scaling regime around $B_{c}$ = 6.73T and its smaller variance of the exponent provide confidence on the scaling form. Remarkably, this is the same universal scaling form and the exponent found in quantum Hall-insulator transitions in bulk two-dimensional electron systems\cite{Sondhi97}.

Although a disorder driven metal-insulator transition in a line junction has been predicted earlier\cite{Renn95,Kane97}, the high quality of the MBE-grown barrier and the momentum conservation in the single particle tunneling lead us to discount the likelihood of disorder playing a prominent role. The ballistic property of edge states further minimizes the possible localization effects associated with disorder. Clear demarcation of the conductance about the quantum critical points and the relative unimportance of disorder lead us to consider the high conductance and the low conductance regimes as a pair of highly correlated ground states.
 Since the two edge states are separated by a tunnel barrier on the order of inter-electron distance, it follows that electron-electron intreaction between the two edge states should play an important role in the electronic transport across the line junction. 

It has been proposed by Kim and Fradkin\cite{Kim03} that inter-edge tunneling in our devices is equivalent to a coupled one-dimensional system interacting through short range interactions. It is postulated that the tunneling between the right- and left-moving edge modes occurs primarily through a weak tunneling center that acts like a point contact. The electron-electron interaction in the model can be accounted through a rigorous mapping of two parallel edge channels into a coupled Luttinger liquid characterized by an effective Luttinger parameter $K$. Depending on the coupling constant between the left and right moving branches, there is a quantum phase transition between a state with no tunneling for $K > 1$ or perfect tunneling $K < 1$. The experimentally observed sequence of critical points represents a series of $K = 1$ quantum critical points between the strongly and weakly tunneling regimes. The sequence of critical point therefore mimics a series of opening and pinching-off of the tunneling center as a function of magnetic field.

While our data is qualitatively consistent with the proposed scenario by Kim and Fradkin, there remains a number of unanswered important questions regarding the physics behind the observed phase transitions, the underlying order on either side of the quantum critical points, and the significance of the similarity to the quantum Hall-insulator transitions. Further theoretical investigation is necessary. 

\section{Plateau Transitions and Inter-Edge Tunneling}

\begin{figure}
\includegraphics[width= \linewidth]{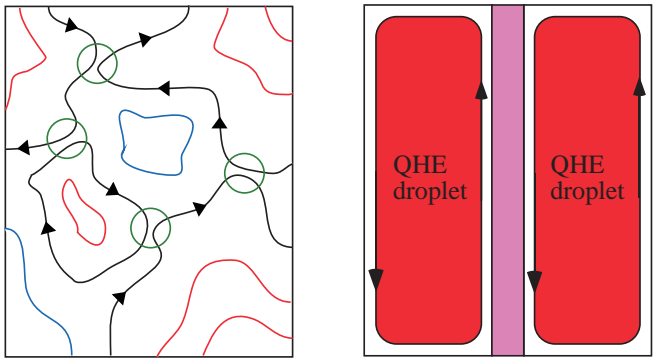}
\caption{\label{Fig:qhe-equi} (a) Contour map of equipotentials of a quantum Hall plateau. Red loops indicate the minima regions and blue loops indicate the maxima regions. Circles indicates tunneling sites where two, counterflowing equipotentials become very close. (b) Tunneling geometry for the present experiment.}
\end{figure}

From the resemblance of the observed transition to a quantum Hall-insulator transition, we consider a scenario that may provide an explanation for the observed critical points and exponents. The transport in the quantum Hall plateau is thought to involve a series of interconnected equipotential ``links'' through a spatially varying random potential. Figure~\ref{Fig:qhe-equi}a illustrates such an equipotential contour for a quantum Hall sample in the Hall plateau. Potential fluctuations produce phase separation of the sample into small quantum Hall droplets with different integral fillings. These droplets are surrounded by an equipotential boundary that define the respective edge state whose size grows and shrinks with the magnetic field. At the plateau transitions, the current flow occurs through tunneling between the adjacent edge states of the quantum Hall droplets at these equipotential links. Such considerations have led to development of the network model of quantum Hall transitions by Chalker and Coddington\cite{Chalker88}.

A mesoscopic realization of coupled quantum Hall droplets is obtained in our samples as illustrated in Fig.~\ref{Fig:qhe-equi}b. The tunneling between small quantum Hall droplets may be realizable as tunneling between the edge states of neighboring quantum droplets separated by a well-defined barrier. In effect, the two-dimensional quantum Hall plateau transitions at small distance may be equivalent to the problem of tunneling between two one-dimensional edge states. The similarity of the scaling form and the exponents, plus the consideration of the respective electronic potential, provides a strong support for such a scenario. Tunneling in line junctions presumably occurs at many places along the barrier unlike the junctures between two quantum Hall droplets which may behave more like a quantum point-contact. Existence of multiple tunneling paths may be responsible for the reduced value of the conductance at the critical points (0.03-0.05$e^{2}/h$). Whether such a conjecture can explain the observed data remains to be explored.

\section{Conclusion}

In conclusion, we have studied the temperature dependent transport 
across a quantum Hall line junction. The tunnel-coupled, 
counterpropagating edge states produce a series of quantum critical 
points between the highly and weakly tunneling regimes.
These critical points indicate a series of quantum 
phase transitions between two correlated one-dimensional ground 
states arising as a result of strong interedge correlation.
Scaling analysis shows that the conductance near the critical 
behavior scales as $|B-B_{c}|T^{-\kappa}, \kappa \approx 0.42$,  
similar to that of quantum Hall-insulator transitions. These results points to presence of strong quantum Hall correlation between the coupled edges of quantum Hall droplets.

\begin{acknowledgments}
 The work at the University of Chicago is supported by
NSF DMR-0203679 and NSF MRSEC Program under DMR-0213745.
\end{acknowledgments}

\end{document}